# Metal-insulator transition in quasi-one-dimensional HfTe$_3$ in the few-chain limit

Scott Meyer[1,2,3,4,†], Thang Pham[1,3,4,5,†], Sehoon Oh[1,4], Peter Ercius[6], Christian Kisielowski[6], Marvin L. Cohen[1,4], and Alex Zettl[1,3,4]*

[1]Department of Physics, University of California at Berkeley, Berkeley, CA 94720, USA

[2]Department of Chemistry, University of California at Berkeley, Berkeley, CA 94720, USA

[3]Kavli Energy NanoSciences Institute at the University of California at Berkeley, Berkeley, CA 94720, USA

[4]Materials Sciences Division, Lawrence Berkeley National Laboratory, Berkeley, CA 94720, USA

[5]Department of Materials Science and Engineering, University of California at Berkeley, Berkeley, CA 94720, USA

[6]The Molecular Foundry, One Cyclotron Road, Berkeley, CA 94720 USA

†These authors contributed equally

*Correspondence to: azettl@berkeley.edu

**ABSTRACT**

The quasi-one-dimensional linear chain compound HfTe$_3$ is experimentally and theoretically explored in the few- to single-chain limit. Confining the material within the hollow core of carbon nanotubes allows isolation of the chains and prevents the rapid oxidation which plagues even bulk HfTe$_3$. High-resolution transmission electron microscopy combined with density functional theory calculations reveals that, once the triple-chain limit is reached, the normally parallel chains spiral about each other, and simultaneously a short-wavelength trigonal anti-prismatic rocking distortion occurs that opens a significant energy gap. This results in a size-driven metal-insulator transition.

Constraining the physical size of solids can dramatically influence their electrical, optical, magnetic, thermal, and mechanical properties. Intrinsically low-dimensional materials, including van der Waals (vdW) bonded quasi-two-dimensional compounds (exemplified by graphite, hexagonal boron nitride, and transition metal dichalcogenides (TMD)) and quasi-one-dimensional compounds (exemplified by transition-metal trichalcogenides (TMT)), are particularly intriguing, in that the bulk state already presents weakened inter-plane or inter-chain bonding, which leads to strong structural, electronic, and phononic anisotropy. [1,2] Constraining the dimensions of these materials down to "atomic thinness" can result in various degrees of additional size quantization with profound consequences.

Recently, the prototypical quasi-one-dimensional TMT conductor $NbSe_3$ was successfully synthesized in the few- to single-chain limit, and unusual torsional wave instabilities were observed. [3] The driving force for the instabilities was proposed to be charging of the chains, which suggests that other TMT compounds with closely related crystal structure might exhibit similar torsional wave instabilities in the few- or single-chain limit.

$HfTe_3$ is an intriguing, but little studied, Group IV TMT with a trigonal prismatic linear chain structure very similar to that of the Group V TMT $NbSe_3$. [4–6] Fig. 1 shows the quasi-one-dimensional crystal structure of $HfTe_3$. Each chain distributes the Te atoms in an isosceles triangle, with the unit cell of $HfTe_3$ containing two trigonal prismatic chains with an inversion center. A characteristic that has inhibited extensive study of HfTe$_3$ is extreme air sensitivity, even for bulk single crystals. [7] Some studies [7,8] suggest that metallic HfTe$_3$ supports a charge density wave (CDW) and possibly filamentary superconductivity, but there are significant discrepancies between reports. Single crystal specimens likely undergo a CDW phase transition at $T_P$=93 K [8], while $T_P$ for polycrystalline specimens is ~80 K. [7,8] Although single crystals have not shown superconductivity down to 50 mK, [8] polycrystalline samples can apparently undergo a superconducting phase transition at $T_c$=1.7 K. [7]

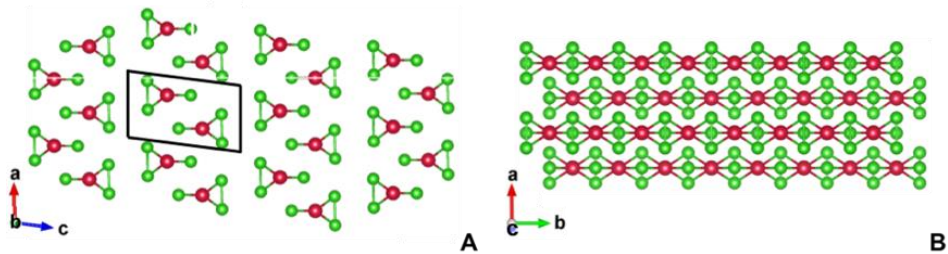

**Figure 1:** Crystal view along the (A) *b* axis and (B) *c* axis, highlighting the quasi-one-dimensional nature of the trigonal prismatic HfTe$_3$ chains, with the unit cell boxed in black. Hf and Te atoms are represented by red and green spheres, respectively.

Here we report the successful synthesis and structural characterization of HfTe$_3$ within the hollow cores of multiwall carbon nanotubes (MWCNT). The selectable inner diameter of the MWCNT constrains the transverse dimension of the encapsulated HfTe$_3$ crystal and thus, depending on the inner diameter of the nanotube, HfTe$_3$ specimens with many chains (~20), down to few chains (3 and 2), and even single isolated chains, are obtained. The MWCNT sheath simultaneously confines the chains, prevents oxidation in an air environment, and facilitates characterization *via* high resolution transmission electron microscopy (TEM) and scanning transmission electron microscopy (STEM). Together with complementary first-principles calculations, we find a coordinated interchain spiraling for triple and double chain HfTe$_3$ specimens, but, in sharp contrast to NbSe$_3$, long-wavelength intrachain torsional instabilities are markedly absent for isolated single chains. Instead, HfTe$_3$ shows a structural transition via a trigonal prismatic rocking distortion to a new, unreported crystal phase, concomitant with a metal-insulator transition, as the number of chains is decreased below four.

HfTe$_3$ is synthesized within carbon nanotubes using a procedure similar to that outlined previously for NbSe$_3$, [3] following HfTe$_3$ growth-temperature protocols. [7] Typically, stoichiometric amounts of powdered Hf along with Te shot (560 mg total), together with 1-4 mg of end-opened MWCNTs, and ~5 mg/cm$^3$ (ampoule volume) of I$_2$ are sealed under vacuum in a quartz ampule and heated in a uniform temperature furnace at 520 °C for 7 days, then cooled to room temperature over 9 days. Energy dispersive spectroscopy (EDS) confirms a 1:3 stoichiometry of encapsulated HfTe$_3$ chains (25.14 at.% Hf, 74.86 at.% Te).

Fig. 2 shows high-resolution TEM images of representative HfTe$_3$ samples encased within MWCNTs, together with simplified side view and cross-sectional view schematics. In Fig. 2A, a 3.85 nm-wide (inner diameter) MWCNT encases ~20 HfTe$_3$ chains (the number of chains is estimated based on the carbon nanotube diameter and a close-packing configuration of the chains). Figs. 2B, 2C and 2D show three, two, and one HfTe$_3$ chain(s) within MWCNTs of inner diameters 2.50 nm, 1.81 nm, and 1.19 nm, respectively. We thus successfully achieve the single-chain limit of HfTe$_3$. Fig. 3 shows STEM images of the few- and single-chain limit of HfTe$_3$ samples encased within MWCNTs, with an atomic model representation of the double- and single-chain limit. Figs. 3A, 3B, and 3C show a triple, double, and single chain of HfTe$_3$ confined within MWCNTs of inner diameters 2.51 nm, 1.69 nm, and 1.21 nm, respectively.

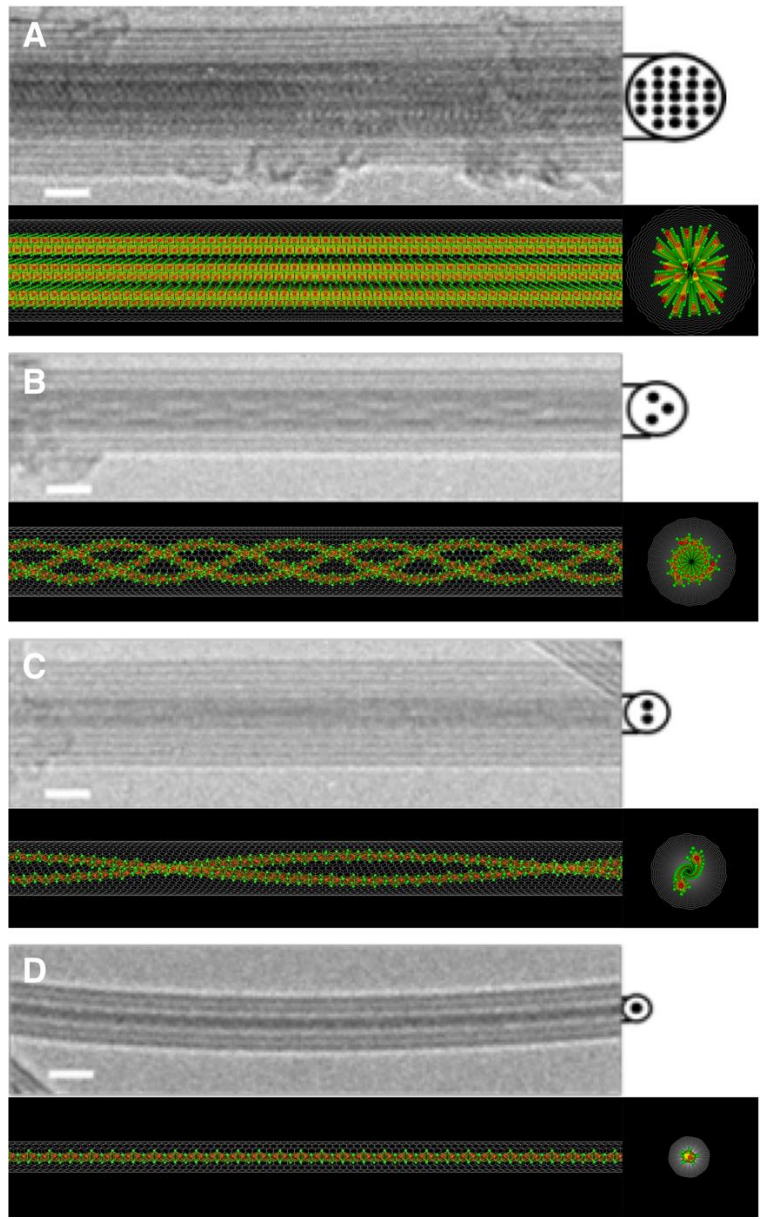

**Figure 2: Encapsulation series from many- to single-chain limit of HfTe₃.** High resolution transmission electron microscopy images of (A) many- (B) triple- (C) double- and (D) single-chain limits of HfTe$_3$ encapsulated within a carbon nanotube. A simplified cross-sectional representation of the filled carbon nanotube is shown to the right of each image, with a model of the chains' filling behavior shown below each image. Scale bars measure 2 nm. All images are underfocused, where Hf and Te atoms appear dark.

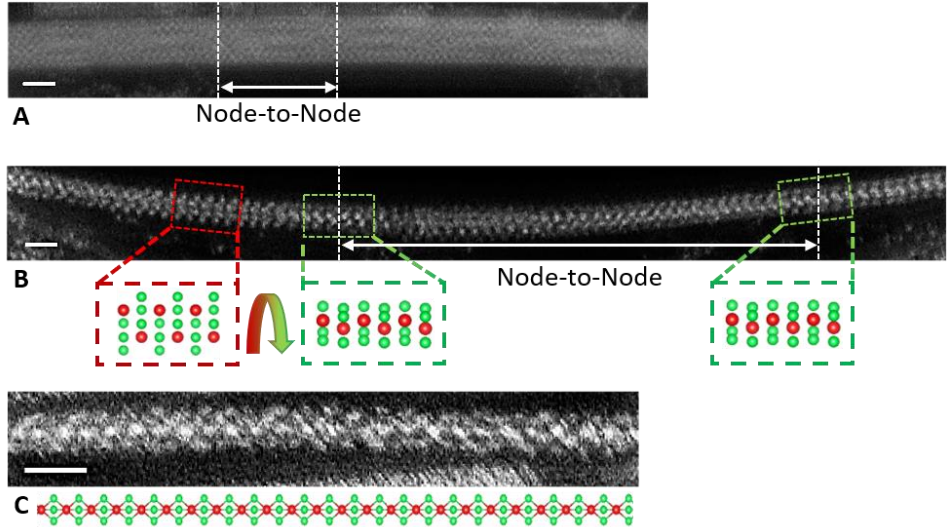

**Figure 3: Encapsulation of single, double, and triple HfTe₃ chains.** Scanning transmission electron microscopy image of (A) triple, (B) double, and (C) single HfTe$_3$ chains encapsulated within a carbon nanotube. Hf and Te atoms appear white in the images. Atomic models below (B) and (C) demonstrate the orientation of the chain(s), where Hf and Te atoms are red and green, respectively. The node-to-node length of the spiraling in (A) and (B) is marked by white dashed lines. Scale bars measure 1 nm.

For the related material NbSe$_3$ in the few-chain limit, 3 or 2 chains spiral around each other in a helical fashion, and in the single chain limit the trigonal prismatic units comprising the chain gradually twist azimuthally as one progresses along the chain axis, comprising a single-chain torsional wave. Figs. 2B, 2C, 3A, and 3B show clearly that HfTe$_3$ displays the same spiraling behavior in the triple- and double-chain limit. For triple HfTe$_3$ chains, the spiraling node-to-node distance ranges from 3.05 to 4.44 nm (Figs. 2B and 3A), while for double HfTe$_3$ chains, the node-to-node distance ranges from 10.60 to 11.07 nm (Figs. 2C and 3B). These observations demonstrate that interchain spiraling, for low chain number, is not unique to NbSe$_3$—it appears to be a general feature of confined TMTs, independent of the chemical composition of the chain. The difference in node-to-node distance of the HfTe$_3$ chains, which is significantly longer when compared to the node-to-node distance for NbSe$_3$ (1.45 to 1.85 nm in triple chain NbSe$_3$, 1.90 to 2.30 nm in double chain NbSe$_3$), is in large part due to the larger tellurium atoms sterically preventing as tight of a spiraling overlap between the chains.

An intriguing question is, does a single chain of HfTe$_3$ encapsulated within a MWCNT support a torsional wave (as does a single chain of NbSe$_3$)? We answer our question by applying high resolution aberration corrected HAADF STEM imaging at 80 kV to encapsulated HfTe$_3$. Fig. 3 shows a STEM image of an encapsulated single chain of HfTe$_3$, along with an atomic model, where the contrast setting does not show the CNT walls. Fig. S1 shows additional encapsulated single-chain HfTe$_3$, along with a higher contrast image to show the CNT walls. No long-wavelength torsional wave is observed in the single-chain limit of HfTe$_3$. Despite common interchain spiraling observed in triple and double chains of both NbSe$_3$ and HfTe$_3$, the single-chain charge-induced torsional wave (CTW) observed for NbSe$_3$ is absent in HfTe$_3$, which points to a fundamental difference between single chains of NbSe$_3$ and HfTe$_3$. In addition, as we show below, the chains themselves in few-to-single-chain specimens of HfTe$_3$ display a completely different kind of structural distortion, that of intracell rocking, which, in sharp contrast to NbSe$_3$, results in a size-driven metal-insulator transition.

To more fully understand the structural distortions of few- to single-chain HfTe$_3$, we perform density functional theory (DFT) calculations. First, we investigate the atomic and electronic structures of a single chain of HfTe$_3$ isolated in vacuum. From the atomic positions of the chains comprising bulk solid, we construct candidate structures using supercells with various length from 1b$_0$ to 12b$_0$ to investigate possible twisting behavior, where b$_0$ is the distance between the nearest Hf atoms. From the constructed candidate structures, atomic structures are optimized by minimizing the total energy. Unexpectedly, all investigated atomic structures of single-chain HfTe$_3$, except for a periodicity $\lambda=1b_0$, show a short-wavelength rocking distortion from a trigonal prismatic (TP) unit cell [Fig. 4C] to a trigonal antiprismatic (TAP) unit cell [Fig. 4H]. This is in sharp contrast to the long-wavelength torsional wave observed in single-chain NbSe$_3$. Figs. 4A-B and 4F-G show the atomic structure and the corresponding electronic structure of single-chain HfTe$_3$ in TP geometry obtained with a periodicity of $\lambda=1b_0$, and rocked TAP geometry with $\lambda=2b_0$. As shown in Fig. 4I-J, the calculated electronic structure of the single-chain indicates a semiconducting transition upon isolation of a single chain, with a significant energy gap of 1.135 eV opening. Additionally, the rocked TAP structure of the HfTe$_3$ chains is observed in chain systems of 3 chains or fewer, leading to a band gap opening, as will be discussed in subsequent sections.

Next, we investigate the atomic and electronic structures of single-chain HfTe$_3$ encapsulated inside a carbon nanotube (CNT). The initial candidate structures of both TP and rocked TAP geometry single chains are constructed using the separately relaxed atomic positions of single-chain HfTe$_3$ isolated in vacuum, and those of an empty (8,8) CNT (indices chosen for convenience). From the candidate structures, the atomic positions of the chain are relaxed by minimizing the total energy, whereas atomic positions of the CNT are fixed. We calculate the binding energy $E_b$ of a single-chain HfTe$_3$, which is defined as $E_b=E_{HfTe3}+E_{CNT}-E_{HfTe3/CNT}$, where $E_{HfTe3}$ is the total energy of the isolated TAP single-chain HfTe$_3$, $E_{CNT}$ is the total energy of an empty CNT isolated in vacuum, and $E_{HfTe3/CNT}$ is the total energy of the joint system of the TP or TAP single-chain HfTe$_3$ encapsulated inside the CNT. The calculated binding energies of TP and TAP single chains are 0.964 and 1.23 eV per HfTe$_3$ formula unit (f.u.), respectively, confirming that the encapsulated single-chain HfTe$_3$ inside CNT also adopts a TAP geometry as in the isolated case. Because of the extremely short wavelength of the rocking TAP distortion, we are unable to resolve it experimentally via (S)TEM.

Figs. 4K-L and 4M-N show the calculated atomic structure of single-chain HfTe$_3$ with TAP geometry encapsulated in the CNT and the corresponding electronic structure. The Fermi energy lies at the energy level of the Dirac point of the CNT, which is inside the gap of the single-chain. As shown in Figs. 4E-F and 4I-J, the states of single-chain HfTe$_3$ near the Fermi energy are not altered appreciably by the confinement, indicating there is no charge transfer between the HfTe$_3$ chain and CNT (unlike the case of encapuslated NbSe$_3$).

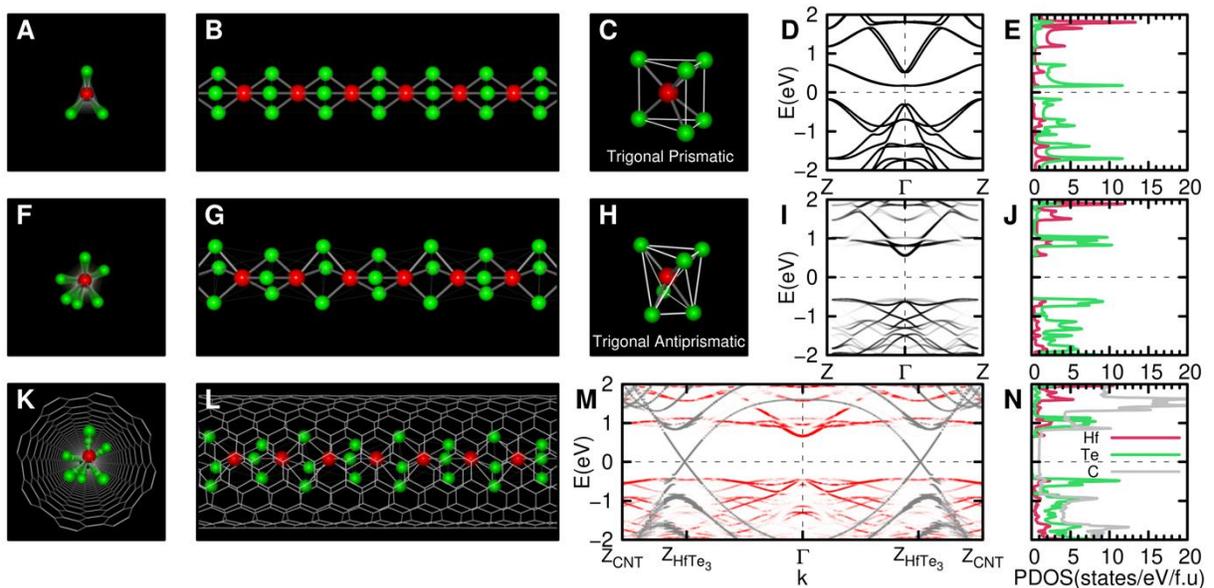

**Figure 4: Calculated atomic and electronic structures of single-chain HfTe$_3$.** The atomic and electronic structures of single-chain HfTe3 isolated in vacuum with (A-E) TP and (F-J) TAP geometry, and (K-N) the TAP single-chain encapsulated inside a (8,8) CNT are presented. In the atomic structure, the red and green spheres represent Hf and Te atoms, respectively. The basic units of (C) the TP and (H) TAP geometry are shown for comparison. In the band structures, the chemical potential is set to zero and marked with a horizontal dashed line. In (M), the bands represented by red and grey lines are projected onto the single-chain HfTe$_3$ and CNT,

respectively, and unfolded with respect to the first Brillouin zone of the unit cell of the single-chain with periodicity λ=$b_0$ and the CNT, where zone boundaries for the chain and CNT are denoted as $Z_{HfTe_3}$ and $Z_{CNT}$, respectively.

The TAP rocking in single-chain $HfTe_3$ vs. the long-wavelength torsional wave instability observed in single-chain $NbSe_3$ is the most notable difference between the two systems. To explore the mechanism dictating such a drastic difference observed at the single-chain limit, two factors are key: (i) the geometry of the unit cell of the chain and (ii) the electronic structure of a single chain in each system.

Because the Te atoms in $HfTe_3$ are distributed as an isosceles triangle in a trigonal prismatic chain, the 3-fold symmetry of the chain is broken and the inversion center of the unit cell is lost when the single-chain limit of $HfTe_3$ is reached. This causes the Te bands near the chemical potential to split. Splitting of the bands reduces the energies of the occupied Te band near the chemical potential and creates a semiconducting gap of 0.341 eV in a single $HfTe_3$ chain, as shown in Figs. S3D-F. However, the total energy of the single chain of $HfTe_3$ can be further lowered by rocking the Te atoms between each Hf metal center into a TAP chain, splitting the Te bands near the chemical potential even more than the TP chain, as shown in Figs. S3G-I. The rocked TAP structure of single-chain $HfTe_3$ has 0.479 eV/f.u. lower total energy than the TP single chain, with the energy gap enlarging from 0.341eV to 1.135 eV in the final rocked TAP structure. We note that we have also investigated an equilateral distribution of the Te atoms, similar to the Se atoms in single-chain $NbSe_3$,[11] shown in Fig. S3A-C, and thereby confirmed that the isosceles distribution in $HfTe_3$ continues to be the energetically preferred structure for all chain numbers.

Splitting of the Te bands in the TAP chain is possible because the single chain of HfTe$_3$ has an even number of electrons in the unit cell. A single chain of NbSe$_3$ has an odd number of electrons in the unit cell, preventing any splitting of the bands, and allowing a metallic band structure with 3-fold symmetry even down to the single-chain limit. Therefore, for single-chain TMTs, we observe either a TP (NbSe$_3$) or TAP (HfTe$_3$) structural arrangement of the chalcogen atoms, depending on elemental composition, leading to metallic or insulating behavior, respectively. The structural difference between NbSe$_3$ and HfTe$_3$ is analogous to the transition-metal dichalcogenides, where some materials (such as MoS$_2$) prefer the trigonal prismatic (1H) structure showing insulating behavior, while others (such as WTe$_2$) prefer the trigonal antiprismatic (1T or 1T') structure showing metallic behavior.

To investigate multi-chain spiraling and possible on-chain rocking of double- and triple-chain HfTe$_3$, we construct several candidate structures isolated in vacuum using the atomic positions of the chains comprising bulk solid with the diameters and periodicities of the spiral wave obtained from experimental evidence and minimize the total energy to determine the fully-relaxed atomic structure. Figs. 5A-C and 5D-F show the relaxed atomic structure, electronic band structure, and projected density of states (PDOS) of the spiraling double- and triple-chain HfTe$_3$, respectively. As shown in Figs. 5A and 5D, the individual chains comprising the spiral double- and triple-chain also rock into the TAP geometry, similar to the single-chain HfTe$_3$, to minimize the total energies of each chain. In turn, each TAP chain spirals around the others in a helical fashion. The obtained electronic structures of spiraling double- [Figs. 5B-C] and triple-chain [Figs. 5E-F] HfTe$_3$ resemble that of the TAP single-chain [Figs. 4I-J]. Spiraling double- and triple-chain HfTe$_3$ has energy gaps of 1.020 and 1.018 eV, respectively, comparable to that of the TAP single-chain, 1.135 eV.

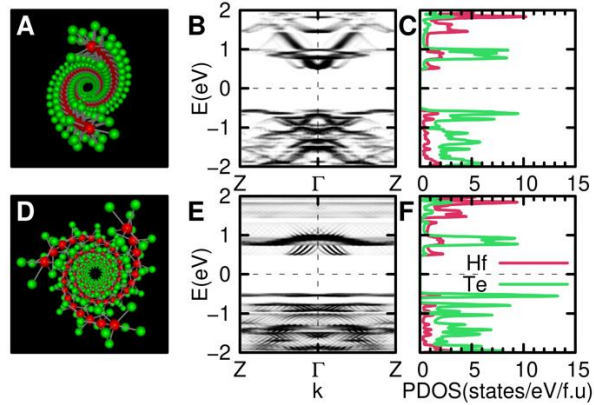

**Figure 5: Calculated atomic and electronic structures of spiraling double, and triple-chain HfTe$_3$.** The atomic and electronic structures of spiral (A-C) double and (D-F) triple chains of HfTe$_3$ isolated in vacuum are presented. In the axial view along the *b*-axis of the unit cell, the red and green spheres represent Hf and Te atoms, respectively. In (B, E) the band structures, the chemical potential is set to zero and marked with a horizontal dashed line and unfolded with respect to the first Brillouin zone of the unit cell of the single-chain with periodicity $\lambda=b_0$, where the Brillouin zone center and the edge are denoted as Γ and Z, respectively. The individual chains comprising triple- and double-chain rock into TAP geometry.

To understand the preferred spiraling pattern and on-chain rocking of double- and triple-chain HfTe$_3$, the competing interactions that exist among free-standing parallel chains and the interactions among encapsulated spiraling chains are analyzed. In bulk down to quadruple chains, strong interchain vdW interactions between the Hf centers and Te atoms on neighboring chains allow for the largest energy stabilization, and this is the largest determining factor in the parallel orientation of the chains. Metallic behavior is maintained from bulk to quadruple chains. Once the triple- and double-chain limit is reached, however, the chains undergo two physical changes. First, the Te ligands rock to form the TAP unit within each chain, which lowers the total chain energy and opens the energy gap. Second, the chains spiral around one another in a helical fashion. Interestingly, spiraling of the double- and triple-chain systems of HfTe$_3$ does not significantly alter the band gap; the rocking distortion into the TAP chain conformation remains the main driving force behind the metal-insulator transition in the few-chain limit of HfTe$_3$.

In summary, on-chain rocking of HfTe$_3$ chains into the TAP geometry drives a metal-insulator transition for chain systems of three or fewer. Quadruple- and higher-chain systems have more neighboring chains with a larger number of interchain vdW interactions between the Hf centers and Te atoms on those neighboring chains, preventing the chains from rocking into the TAP geometry, which maintains the metallic behavior. Encapsulation of the triple- and double-chain limit within a CNT promotes the spiraling of the chains. The spiraling enhances the vdW interactions between the chains and the CNT inner wall and further stabilizes the chains.

**ACKNOWLEDGEMENTS**

**Funding**: This work is primarily funded by the U.S. Department of Energy (DOE) Office of Science, Office of Basic Energy Sciences, Materials Sciences and Engineering Division, under contract DE-AC02-05-CH11231 within the sp$^2$-Bonded Materials Program (KC2207), which provided for synthesis of the chains, TEM structural characterization, and theoretical modeling and electronic energy band calculations of the few- and single-chain limits of HfTe$_3$. The elemental mapping work was funded by the DOE Office of Science, Office of Basic Energy Sciences, Materials Sciences and Engineering Division, under contract DE-AC02-05-CH11231 within the van der Waals Heterostructures Program (KCWF16). Work at the Molecular Foundry (TEAM 0.5 characterization) was supported by the DOE Office of Science, Office of Basic Energy Sciences, under contract DE-AC02-05-CH11231. Support was also provided by NSF grants DMR-1206512 (which provided for preparation of opened nanotubes) and DMR1508412 (which provided for theoretical calculations of uncharged TMT materials). **Author Contributions:** S.M., T.P., and A.Z. conceived the idea; S.M. synthesized the materials; S.M., T.P., P.E., and C.K. conducted TEM studies; S.O. performed DFT calculations; A.Z. and M.L.C. supervised the project; and all authors contributed to the discussion of the results and writing of the manuscript. **Competing Interests:** Authors have no competing interests.

Supplementary Materials

# Metal-insulator transition in quasi-one-dimensional HfTe$_3$ in the few-chain limit

Scott Meyer[1,2,3,4,†], Thang Pham[1,3,4,5,†], Sehoon Oh[1,4], Peter Ercius[6], Christian Kisielowski[6], Marvin L. Cohen[1,4], and Alex Zettl[1,3,4]*

[1]Department of Physics, University of California at Berkeley, Berkeley, CA 94720, USA

[2]Department of Chemistry, University of California at Berkeley, Berkeley, CA 94720, USA

[3]Kavli Energy NanoSciences Institute at the University of California at Berkeley, Berkeley, CA 94720, USA

[4]Materials Sciences Division, Lawrence Berkeley National Laboratory, Berkeley, CA 94720, USA

[5]Department of Materials Science and Engineering, University of California at Berkeley, Berkeley, CA 94720, USA

[6]The Molecular Foundry, One Cyclotron Road, Berkeley, CA 94720 USA

†These authors contributed equally

*Correspondence to: azettl@berkeley.edu

Materials

Carbon nanotubes are purchased from CheapTubes (90% SW-DW CNTs), Hafnium powder from Fisher Scientific (99.6%, 325 mesh), Tellurium shot from Sigma-Aldrich (99.999%, 1-2mm), and Iodine from Spectrum Chemicals (99.8%, resublimed).

Transmission Electron Microscopy (TEM) Characterization

After synthesis, filled carbon nanotubes are cast onto lacey carbon TEM grids for characterization. A JEOL 2010 microscope (80 kV) is used for high-resolution imaging (HR-TEM), Titan-X (80 kV) for energy dispersive spectroscopy (EDS), and TEAM 0.5 (aberration corrected, 80 kV) for scanning transmission electron microscopy (STEM) imaging.

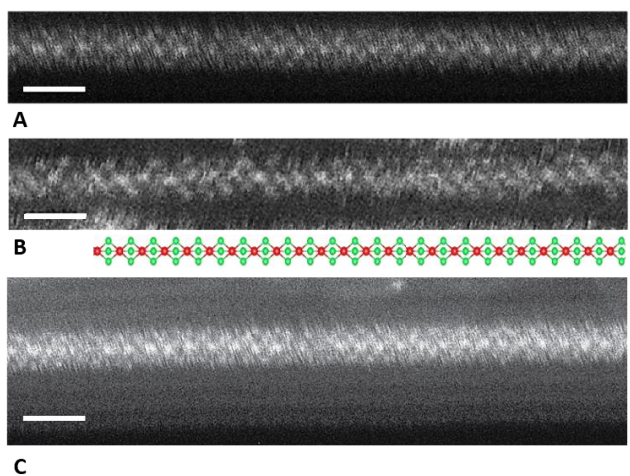

**Fig. S1: Encapsulation of single HfTe$_3$ chains.** Scanning transmission electron microscopy image of (A-C) single HfTe$_3$ chains encapsulated within a carbon nanotube. (A) and (C) show the same single-chain HfTe$_3$, however (C) is shown in high contrast where the CNT walls are visible. Hf and Te atoms appear white in the images. An atomic model below demonstrates the orientation of the chain in both A and B, where Hf and Te atoms are red and green, respectively. Scale bars measure 1 nm.

Calculation Methods

We perform first-principles calculation based on density functional theory(DFT). We use the generalized gradient approximation, [9] norm-conserving pseudopotentials, [10] and localized pseudo-atomic orbitals for the wavefunction expansion as implemented in the SIESTA code. [11] The spin-orbit interaction is considered using fully relativistic j-dependent pseudopotentials [12] in the l-dependent fully-separable nonlocal form using additional Kleinman-Bylander-type projectors. [13,14] We use 1×512×1 Monkhorst-Pack k-point mesh for finite chains, and 40×64×24 for bulk. Real-space mesh cut-off of 1000 Ry is used for all of our calculations. The van der Waals interaction is evaluated using the DFT-D2 correction. [15] For finite chains, a vacuum region of 50 Å × 50 Å perpendicular to the chain is used and dipole corrections are included to reduce the fictitious interactions between chains generated by the periodic boundary condition in our supercell approach. [16]

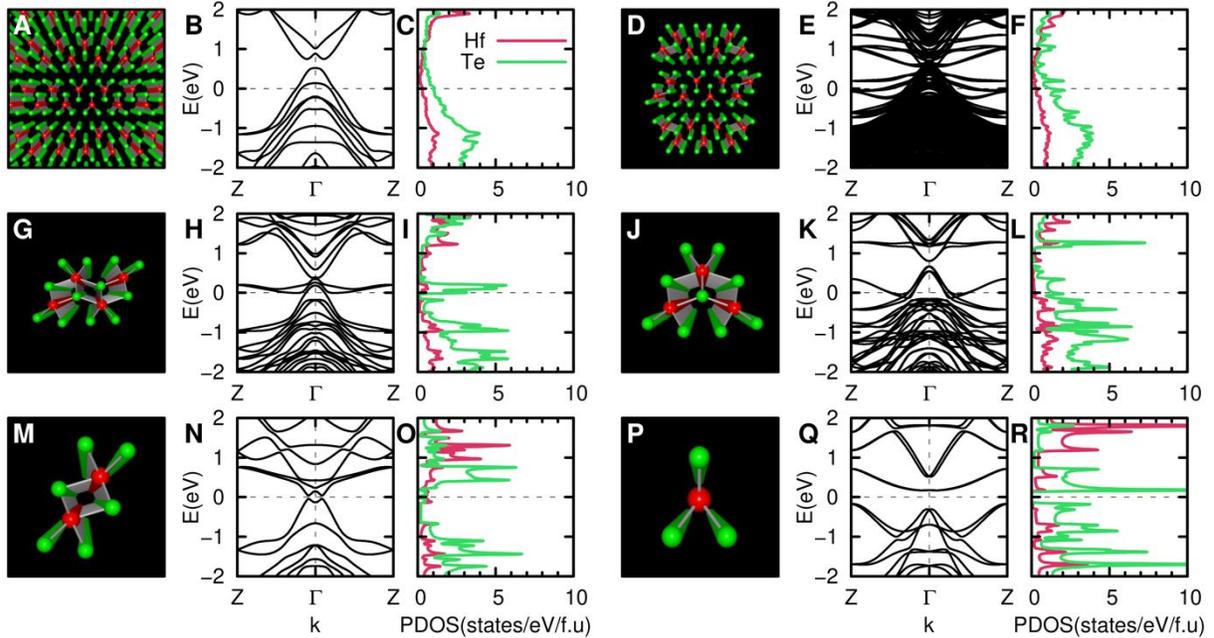

**Figure S2: Calcuated atomic and electronic structures of HfTe$_3$ bulk and finite parallel chains.** The atomic and electronic structures of HfTe$_3$ bulk crystal and parallel finite chains with TP geometry isolated in vacuum are presented. (A-C) The bulk crystal, (D-F) 22-chain, (G-I) quadruple-chain, (J-L) triple-chain, (M-O) double-chain, and (P-R) single-chain. In the axial view along the *b*-axis of the unit cell, the red and green spheres represent Hf and Te atoms, respectively. In the band structures, the chemical potential is set to zero and marked with a horizontal dashed line and the Brillouin zone center and the edge are denoted as Γ and Z, respectively.

To compare the stability of isosceles vs. equilateral distribution of the Te atoms in HfTe$_3$, we also investigate the atomic and electronic structure of the TP single-chain HfTe$_3$ with Te atoms distributed in an equilateral triangle, as shown in Fig. S3A-C. Calculation results from the TP single-chain HfTe$_3$ with equilateral distribution of the Te atoms indicates that the chain is metallic and has 0.193 and 0.672 eV/f.u. higher total energy than the semiconducting isosceles distribution of Te atoms in TP and TAP single-chain HfTe$_3$, respectively.

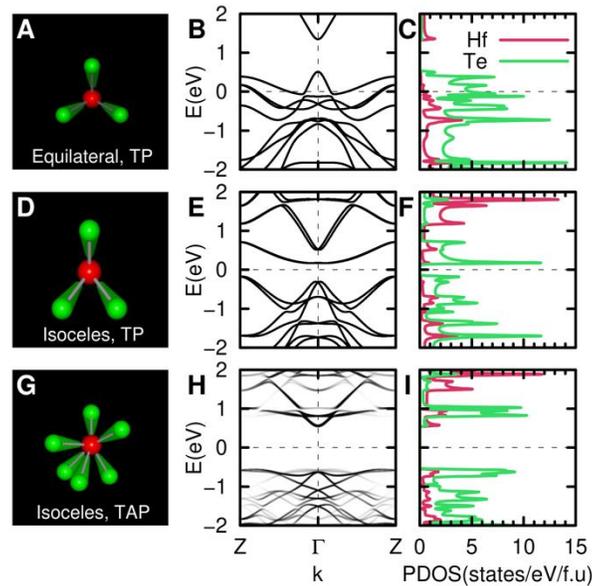

**Figure S3: Calcuated atomic and electronic structures of TP and TAP geometry single-chain HfTe$_3$ with Te atoms forming equilateral and isoceles triangles.** The atomic and electronic structures of single-chain HfTe$_3$ isolated in vacuum are presented. (A-C) The TP single-chain HfTe$_3$ with an equilateral triangle distribution of Te atoms, (D-F) the TP and (G-I) TAP single-chain HfTe3 with an isoceles triangle distribution of Te atoms. In the axial view along the *b*-axis of the unit cell, the red and green spheres represent Hf and Te, respectively. In the band structures, the chemical potential is set to zero and marked with a horizontal dashed line and the Brillouin zone center and the edge are denoted as Γ and Z, respectively. In (H), the bands

are unfolded with respect to the first Brillouin zone of the unit cell of the single-chain with length of $b_0$.

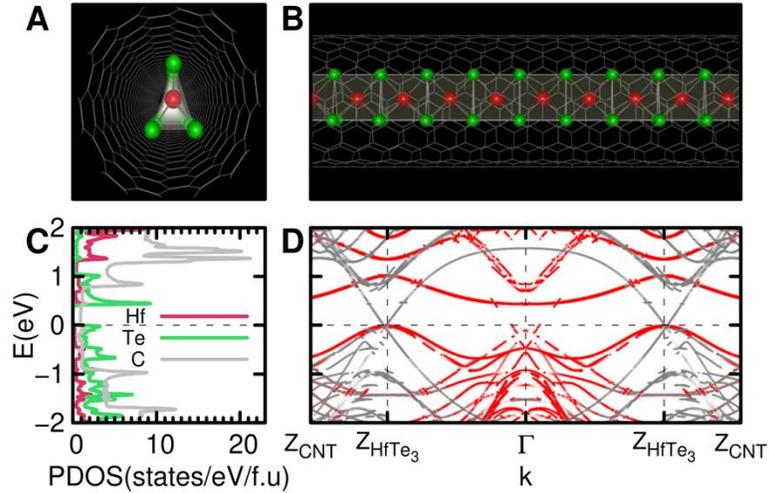

**Figure S4: Calcuated atomic and electronic structures of TP geometry single-chain HfTe$_3$ encapsulated a CNT.** (A-B) The atomic and (C-D) electronic structures of TP geometry single-chain HfTe$_3$ encapsulated in an (8,8) CNT are presented. The atomic structure is obtained with a constraint to force the TP symmetry in the atomic structure optimization. In (C,D), the Fermi energy is set to zero and marked with a horizontal dashed line. In (D), the bands represented by red and grey line are projected onto the chain and CNT, respectively, and unfolded with respect to the first Brillouin zone of the unit cells of the single-chain and CNT, where zone boundaries for the chain and CNT are denoted as $Z_{HfTe_3}$ and $Z_{CNT}$, respectively.